\documentclass[slac_one]{revtex4}
\usepackage{graphicx}
\usepackage{fancyhdr}
\begin{document}

\title{Direct Photons at RHIC} 

%

\author{G. David (for the PHENIX Collaboration)}
\affiliation{BNL, Upton, NY 11973, USA}

\begin{abstract}
Direct photons are ideal tools to investigate kinematical and
thermodynamical conditions of heavy ion collisions since they are emitted
from all stages of the collision and once produced they leave the
interaction region without further modification by the medium.
The PHENIX experiment at RHIC has measured direct photon production in
p+p and Au+Au collisions at 200 GeV over a wide transverse
momentum ($p_T$) range. The $p$+$p$ measurements
allow a fundamental test of QCD, and serve as a baseline when we
try to disentangle more complex mechanisms producing high $p_T$
direct photons in Au+Au.  As for thermal photons in Au+Au we overcome the
difficulties due to the large background from hadronic decays by measuring
"almost real" virtual photons which appear as low invariant mass $e^+e^-$
pairs: a significant excess of direct photons is measured above the above
next-to-leading order perturbative quantum chromodynamics calculations.
Additional insights on the origin of direct photons can be gained with the
study of the azimuthal anisotropy which benefits from the increased
statistics and reaction plane resolution achieved in RHIC Year-7 data.

\end{abstract}

\maketitle

\thispagestyle{fancy}


\section{INTRODUCTION} 

Direct photons - defined as those not originating from (final state)
hadron decays - are an important tool with unique capabilities to
study both elementary particle and heavy ion collisions.  By not
interacting strongly they have a large mean free path and leave the
interaction region unaltered even if they propagate through the hot,
dense medium assumed to be formed in relativistic heavy ion
collisions.  The lowest order QCD processes generating photons -
quark-antiquark annihilation ($q\bar{q}\rightarrow g\gamma$) 
and quark-gluon Compton scattering ($qg\rightarrow q\gamma$) - 
are reasonably well understood~\cite{aurenche2006} thus photons are a
sensitive probe of
modifications of the parton distribution functions in nuclei (nPDFs).
Furthermore, in $p$+$p$ collisions at higher transverse momenta ($p_T$) 
the Compton process dominates, thus the photon ``calibrates'' the
total energy of the opposing gluon jet (apart of the uncertainty 
stemming from the intrinsic $k_T$ of partons).  Setting the energy
scale of the hard scattered parton (and the ensuing jet) is even 
more important in heavy ion collisions, where the partons lose a 
significant fraction of their energy while crossing the
medium~\cite{ppg079}. 

At higher order fragmentation (or Bremsstrahlung) photons contribute
both in $p$+$p$ and heavy ion collisions; the calculated rates in $p$+$p$ are
consistent with recent measurements but the uncertainties are large
both on the theoretical and experimental side~\cite{ppg060}.
It is very important to understand all $p$+$p$ processes as well as
possible because in heavy ion collisions several new photon production 
mechanisms emerge or become significant, and often it is exactly the
{\it difference} from $p$+$p$ behavior that reveals the new physics.
Case in point: the nuclear modification factor $R_{AA}$, the ratio of
a cross-section measured in A+A collisions and the corresponding $p$+$p$
cross-section scaled by the nuclear thickness function.  If $R_{AA}$
is different from unity, it invariably signals some new physics in A+A
with respect to $p$+$p$ collisions.  If $R_{AA}=1$ the interpretation is
somewhat less clear: either the physics mechanisms producing photons
are exactly the same as in $p$+$p$, or there might possibly be a
``conspiracy'' of mechanisms, some increasing, others decreasing
$R_{AA}$ but ultimately balancing each other.  Also, one has to be
careful when interpreting direct photon $R_{AA}$.  For high $p_T$
hadrons (thought to be mostly leading fragments of jets) scaling the
$p$+$p$ cross-sections with the nuclear thickness to obtain the expected
A+A rates (colloquially ``scaling with the number of binary
nucleon-nucleon collisions'' or ``$N_{coll}$ scaling'') 
is legitimate due to the isospin
symmetry of protons and neutrons in the nucleus.  The same is not true
for direct photons: due to the electromagnetic coupling the photon
cross section is proportional to the sum $\Sigma e_q^2$ of quark
charges squared, different for $p$ and $n$, causing a trivial
deviation of $R_{AA}$ from unity (``isospin effect''~\cite{arleo2006})
which may either mask or enhance other effects changing $R_{AA}$.
In fact, the cleanest way to generate direct photon $R_{AA}$ would be
to compare to a properly scaled mixture of $pp$, $pn$ and $nn$ 
cross-sections at the same energy, which could (and in the author's
private opinion, should) be measured at RHIC 
from $d$+$d$ collisions, since the above three reactions could be tagged
event-by-event. 

While direct photons carry a plethora of (almost) unbiased
information, measuring them is very challenging due to the large
background from hadron decays.  Substantial statistics and a thorough
understanding of systematic errors are required.  If in addition to
the total direct photon yield one wants to {\it disentangle} the
contributions from different physics sources (with minimal theoretical
input and assumptions) the difficulties are even higher - but not
insurmountable.  Studying the emergence and evolution of new phenomena
and at the same time tight control of the systematics is crucial.  In 
specific, measuring various colliding systems (from $p$+$p$ to the
heaviest A+A) at different c.m.s. energies in the very same
experiment goes a long way toward this goal.  The flexibility of the
accelerator and of the experiments at RHIC offers such an opportunity.

\section{DIRECT PHOTONS AT RHIC - HIGH $p_T$ AND THE THERMAL REGION}

Direct photon cross-sections in $\sqrt{s}=200$GeV $p$+$p$ collisions at
RHIC were first published in the limited $5<p_T<8$GeV/$c$ range
in~\cite{ppg049}, followed by a measurement in the $3<p_T<16$GeV/$c$
range with considerably smaller systematic errors and detailed
comparison to NLO pQCD calculations~\cite{ppg060}.  It was found that
above $p_T>5$GeV/$c$ (where the systematic errors of the experiment
and theory were comparable) the data are well described by the theory,
including the fraction of isolated photons.  This was welcome news
after the ``controversial situation''~\cite{aurenche2006} of the past
decade.  Since RHIC collides polarized protons, a measurement of the
polarized gluon structure functions is now within reach.  The data
also provide a measured - rather than calculated - baseline for the
nuclear modification in heavy ion collisions ({\it modulo} the
isospin-effect).

The first direct photon invariant yields up to $p_T=13$GeV/$c$ 
from $\sqrt{s_{NN}}=200$GeV Au+Au    
collisions were published in~\cite{ppg042}, covering all collision
centralities (impact parameter ranges) and in the same paper the
direct photon $R_{AA}$, using an NLO pQCD reference and 
integrated above $p_T>5$GeV/$c$ was shown to be unity within errors 
for all centralities.  This was a landmark result because it
retroactively validated the concept of ``$N_{coll}$-scaling'' and
$R_{AA}$ itself in studying ``jet-quenching'' at high $p_T$ 
with single inclusive hadron spectra~\cite{ppg003,ppg048}.
Although recent developments in theory and better data help to draw a
more nuanced picture of how, where and when high $p_T$ photons are
produced (in addition to primordial hard scattering), the above
conclusion is still sane, not the least thanks to the large
differences between the effects: $\pi^0$-s are suppressed by a factor of 5,
while direct photons (still {\it mostly} from hard scattering) are
suppressed much less, or not at all.

\begin{figure*}[t]
\centering
\includegraphics[width=115mm]{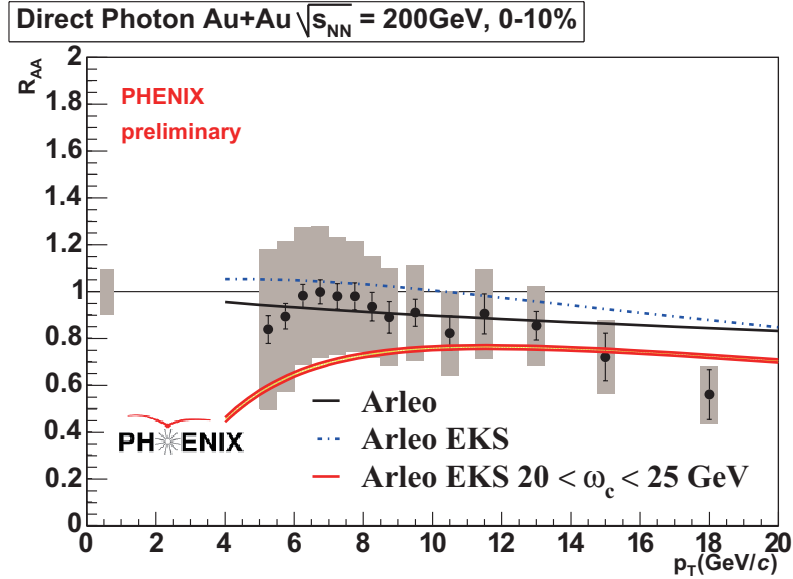}
\caption{Direct photon $R_{AA}$ in central (0-10\%) 
$\sqrt{s_{NN}}=200$GeV Au+Au  collisions.  The denominator is a fit to
the PHENIX preliminary Run-5 $p$+$p$ data.  The theoretical curves are LO
calculations from~\cite{arleo2006}.  The solid line shows the pure
isospin effect, the dash-dotted line shows the combined effect of isospin
and (anti)shadowing (EKS), finally the band at the bottom combines
isospin and antishadowing with photon quenching. }\label{fig:raa}
\end{figure*}

More recent (and still preliminary) data on direct photon $R_{AA}$ and
comparisons to theoretical calculations are shown on
Fig.~\ref{fig:raa}.  With the $p_T$ range now extended
to 20GeV/$c$ photons appear to be somewhat suppressed.  As
mentioned before, some of this might be of trivial origin; the solid 
line predicts a 15\% drop due to the isospin effect alone.  
Counteracting to this - particularly at medium $p_T$, {\it i.e.} in the
$0.1<x=2p_T/\sqrt{s}<0.2$ region - is the anti-shadowing (dash-dotted
curve).  On the other hand, if parton energy loss is added, 
direct photon production is suppressed in the entire range (band at
the bottom of the figure).

While the energy loss in the medium suppresses the yield of high $p_T$
particles, possibly including photons, the ``jet-photon conversion''
mechanism~\cite{fries2003} may increase high $p_T$ photon production.  
In this process a hard scattered quark (proto-jet) interacts with a
gluon or antiquark of the medium.  The collision has a collinear
singularity thus the outcoming photon carries the full momentum of the
original parton.  Note that since antiquarks abund in the thermalized
medium, $q\bar{q}$ annihilation is no longer suppressed with respect
to Compton-scattering of gluons thus a new channel to direct photon
production opens up.

An up-to-date overview of the known/assumed direct photon production
mechanisms in heavy ion collisions is given in~\cite{turbide2008} and
detailed, centrality-dependent calculations are shown
in~\cite{liu2008}.  The tantalizing question is whether the
contributions from the individual components can be disentangled
experimentally?  In other words: can the theory be tested?  
The answer is a tentative
``yes'' with a possible path laid out in~\cite{david2008} and briefly
repeated here.  

First, one has to realize that future, substantially larger data
volumes not only decrease the statistical errors, but they also help
to apply more sophisticated analysis techniques that reduce the
systematic errors.  Also, they allow meaningful analyses of
multiple-differential quantities like azimuthal asymmetries of photon
production.  While such asymmetries are not expected from the
primordial hard scattering, medium-related photon sources will exhibit
them.  By now it is well established that due to the azimuthally
asymmetric overlap geometry in all but the most central
nucleus-nucleus collisions the pressure gradients and thus the
momentum distributions of final state hadrons are asymmetric in 
azimuth (``flow'').  The second Fourier-component ($v_2$) of the
azimuthal distribution of momenta is positive for all hadrons,
therefore, $v_2>0$ for all hadron decay photons as well (the
background to the direct photon measurement).  Also, $v_2>0$ for jet
fragmentation photons.  On the other hand medium-induced photons
typically exhibit $v_2<0$, corresponding to the longer average
pathlength of the parton in the medium.  However, within this category
jet-photon conversion photons are isolated whereas Bremsstrahlung
photons are not.  These properties might help to disentangle the
different production mechanisms of direct photons.

\begin{figure*}[t]
\centering
\includegraphics[width=115mm]{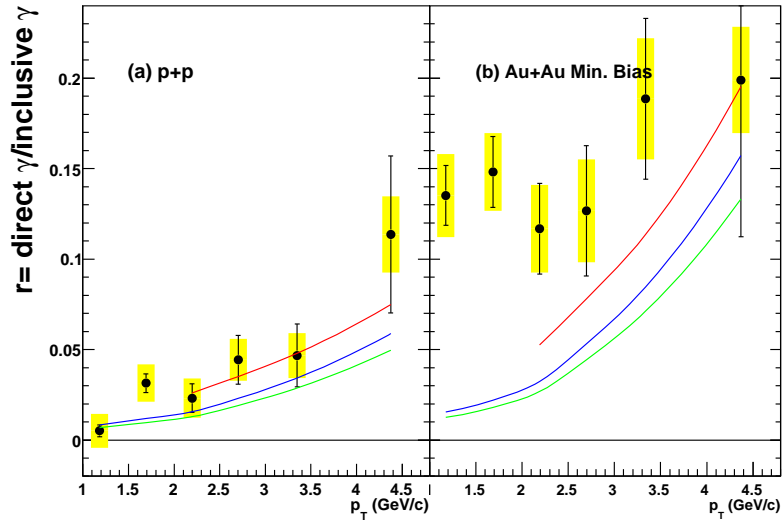}
\caption{The fraction of the direct photon component as
a function of $p_{\rm T}$ in (a) $p$+$p$ and (b) Au+Au (min. bias).
The error bars and the error band represent the statistical and
systematic uncertainties, respectively. The curves are
from a NLO pQCD calculation.} \label{fig:thermal}
\end{figure*}

Thermal radiation of photons from the quark-gluon plasma is expected
to dominate in the $1<p_T<3$GeV/$c$ region, where the measurement of
(real) direct photons with classic calorimetry is nearly impossible
due to the large hadron decay background.  However, one can measure
``quasi-real'' virtual photons which manifest themselves as low mass
$e^+e^-$ pairs~\cite{ppg086}.  With this technique - applied for the
first time in heavy ion collisions - the PHENIX experiment measured
the fraction of direct photons in the total inclusive photon spectrum
both in $p$+$p$ and Au+Au collisions.  The results are shown on
Fig.~\ref{fig:thermal} and compared to NLO pQCD calculations.  In
$p$+$p$ the results are clearly consistent with the calculations (no
new, unknown sources), while in Au+Au we observe a statistically
significant, 10\% excess at low $p_T$.  Converting this excess into
invariant yield and fitting it with an exponential we
find~\cite{ppg086} an inverse slope $T=221\pm23(stat)\pm18(sys)$MeV.
Note that lattice QCD predicts a phase transition already at
$\sim$170MeV, far below the value fitted to our data.
Also, since the observed direct photon spectrum is the integral over
the entire thermal history of the system, the above $T\sim220$MeV
estimate is almost certainly a lower bound only, and the initial
temperature $T_i$ of the system is much higher.

\begin{acknowledgments}
This work was supported by the Office of Science of the 
Department of Energy.
\end{acknowledgments}



\begin{thebibliography}{99} 
\bibitem{aurenche2006}
P. Aurenche {\it et al.}, ``Recent critical study of photon production
in hadronic collisions'', Phys. Rev. D {\bf 73}, 094007 (2006)

\bibitem{ppg079}
A. Adare {\it et al.}, ``Quantitative constraints on the transport
properties of hot partonic matter from semi-inclusive single high
$p_T$ pion suppression in Au+Au collisions at
$\sqrt{s_{NN}}$=200GeV'',  Phys. Rev. C {\bf 77}, 064907 (2008)

\bibitem{ppg060}
S.S. Adler {\it et al.}, ``Measurement of Direct Photon Production in
$p$+$p$ Collisions at $\sqrt{s}=200GeV$'', 
Phys. Rev. Lett. {\bf 98}, 012002 (2007)

\bibitem{arleo2006}
F. Arleo, ``Hard pion and prompt photon at RHIC, from single to double
inclusive production'', JHEP09 (2006) 015

\bibitem{ppg049}
S.S. Adler {\it et al.}, ``Midrapidity direct photon production in $p$+$p$
collisions at $\sqrt{s}=200$GeV'', Phys. Rev. D {\bf 71} 071102(R) (2005)

\bibitem{ppg042}
S.S. Adler {\it et al.}, ``Centrality Dependence of Direct Photon
Production in $\sqrt{s_{NN}}=200$GeV Au+Au Collisions'', 
Phys. Rev. Lett. {\bf 94}, 232301 (2005)

\bibitem{ppg003}
K. Adcox {\it et al.},
``Suppression of Hadrons with Large Transverse Momentum in Central
Au+Au Collisions at $\sqrt{s_{NN}}=130$GeV'',
Phys. Rev. Lett. {\bf 88}, 022301 (2002)

\bibitem{ppg048}
K. Adcox {\it et al},
``Formation of dense partonic matter in relativistic nucleus-nucleus
collisions at RHIC'',
Nucl. Phys. {\bf A757}, (2005) 184

\bibitem{fries2003}
R.J. Fries, B. Muller, D.K. Srivastava,
``High-Energy Photons from Passage of Jets through Quark-Gluon
Plasma'', 
Phys. Rev. Lett. {\bf 90}, 132301 (2003)

\bibitem{turbide2008}
S. Turbide, C. Gale, E. Frodermann, U. Heinz,
``Electromagnetic radiation from nuclear collisions at
ultrarelativistic energies'',
Phys. Rev. C {\bf 77}, 024909 (2008)

\bibitem{liu2008}
F.-M. Liu, T. Hirano, K. Werner, Y. Zhu,
``Centrality-dependent Direct Photon $p_t$ spectra in Au+Au Collisions
at RHIC'', arXiv:0807.4771

\bibitem{david2008}
G. David, R. Rapp, Z. Xu,
``Electromagnetic probes at RHIC-II'', 
Phys. Rep. {\bf 462} (2008) 176

\bibitem{ppg086}
A. Adare {\it et al.},
``Enhanced production of direct photons in Au+Au collisions at
$\sqrt{S_{NN}}$=200 GeV'',
arXiv:0804.4168


\end{thebibliography}
\end{document}